\documentclass[conference]{IEEEtran}
\IEEEoverridecommandlockouts    % to override locked commands

\usepackage{multirow}
\usepackage{lipsum}
\usepackage{amsmath}
\usepackage{amssymb}
\usepackage[ruled,vlined]{algorithm2e}
\usepackage{graphicx}
\usepackage{comment}
\graphicspath{{./Figures/}}

\usepackage{xcolor}
\usepackage{subcaption} 

\usepackage{adjustbox}

\definecolor{DarkRed}{RGB}{180,30,40}
\definecolor{DeepBlue}{RGB}{0,70,160}
\definecolor{DarkGreen}{RGB}{0,130,70}
\definecolor{Violet}{RGB}{120,0,140}
\definecolor{BurntOrange}{RGB}{200,100,0}
\definecolor{TealBlue}{RGB}{0,120,140}

\usepackage{tikz}
\usetikzlibrary{positioning, arrows.meta, fit, calc, backgrounds, scopes}

\tikzset{
    bus/.style={draw, line width=0.5pt, double, double distance=.75pt,},
    busarrow/.style={bus,-{Latex[length=2mm]}},
    block/.style={draw, inner sep=.5em, align=center, fill=white}
}

\newlength{\ieeecolw}
\newlength{\horsep}
\newlength{\versep}
\newcommand{\udelta}{{u}_{\delta_c}}
\newcommand{\ufxr}{{u}_{\tau_c}}

\title{\LARGE \bf A Model Predictive Control Framework for Assisted Vehicle Drifting}

\author{
	\parbox{\textwidth}{%
		\centering
		Marco Cortese$^{1}$, Antonio Gallina$^{2}$, Matteo Grandin$^{2,3}$, Giovanni Righetti$^{1}$, Mattia Bruschetta$^{2}$, Basilio Lenzo$^{1}$%
	}%
	\thanks{$^{1}$Department of Industrial Engineering, Università degli studi di Padova, Padova, Italy
		{\tt\small basilio.lenzo@unipd.it}}%
	\thanks{$^{2}$Department of Information Engineering, Università degli studi di Padova, Padova, Italy
		}%
    \thanks{$^{3}$CRF, Università degli studi di Padova, Padova, Italy}%    
}
\begin{document}
	
	\maketitle
	\thispagestyle{empty}
	\pagestyle{empty}
	
	%%%%%%%%%%%%%%%%%%%%%%%%%%%%%%%%%%%%%%%%%%%%%%%%%%%%%%%%%%%%%%%%%%
	\begin{abstract}
		Model Predictive Control (MPC) has been widely applied to autonomous vehicle drifting. Assisted drifting, that is where the driver remains in the loop, is still comparatively underexplored. Existing approaches often rely on restrictive assumptions, such as precomputed drift equilibria, full actuation authority, or prior path knowledge, which limit applicability to expert drivers. This paper proposes a nonlinear model predictive control (NMPC) framework for assisted drifting on a rear-wheel-drive vehicle. Through steer-by-wire and drive-by-wire interfaces, the controller decouples driver commands from direct actuator inputs, allowing the driver to regulate the desired sideslip through the steering wheel while the NMPC maintains vehicle stability. A dedicated activation logic ensures that the controller engages only under deliberate driver intent. High-fidelity simulations show that the proposed architecture can stabilize drifting maneuvers using a simple single-track prediction model with basic tire dynamics, even when the sideslip reference is continuously varied by the driver.
            % The approach enables a transparent, human-in-the-loop drift assistance system within realistic actuation limits.
	\end{abstract}

	
	%%%%%%%%%%%%%%%%%%%%%%%%%%%%%%%%%%%%%%%%%%%%%%%%%%%%%%%%%%%%%%%%%%
	\section{Introduction}
	\label{sec:introduction}
    % \begin{itemize}
    %     \item Drifting motivation
    %     \item Autonomous drifting (techniques: inverted model, LQR, fuzzy PID, MPC)
    %     \item Assisted drifting (motivation fun to drive guardare paper Velenis)
    %     \item Contributions: eq. drift online, simple control model, generalizable (throttle not torque), VI-CRT validation, logica attivazione.
    % \end{itemize}

Vehicle drifting, sustained high-sideslip cornering with rear-tire saturation and counter-steer, sits at the intersection of three distinct engineering motivations \cite{27Abdulrahim2006}. In motorsport, expert drivers deliberately use this technique to increase the range of achievable vehicle trajectories on low-grip surfaces, such as those typical of rally stages. In the active safety domain, operating at the handling limit may be the only viable option in certain emergency avoidance scenarios where a conventional controller is provably insufficient \cite{18Zhao2022}. Finally, the advent of torque-vectoring drivetrains, particularly in multi-motor electric vehicles, has simplified the automation of drifting maneuvers. Per-wheel torque control provides actuation bandwidth and precision that were previously unavailable, overcoming the mechanical limitations that had made drift control difficult to implement in an automated fashion \cite{25Lenzo2024}.

The theoretical foundation for model-based drift control lies in the analysis of drift equilibria. Hindiyeh and Gerdes showed that a three-state bicycle model admits drift operating points that are open-loop unstable saddle points %: the front tire is approximately saturated in lateral force, leaving rear drive force and steering as the effective handles for yaw rate and sideslip, respectively 
\cite{4Hindiyeh2009}. %This structural insight has motivated 
A progression of increasingly capable controllers followed, from successive-loop model-inversion architectures \cite{2Velenis2011,6Hindiyeh2014} to general curved-path tracking \cite{7Goh2019} and LQR designs exploiting the near-linear behaviour around drift equilibria \cite{8Peterson2023}, with the unifying limitation that they all rely on a pre-specified operating point and lack a principled mechanism to simultaneously enforce actuator bounds, safety margins, and tracking objectives over a predictive horizon. In this setting, the control task is more naturally formulated as a constrained optimal control problem, in which steering and rear-drive inputs must be selected to achieve the desired vehicle evolution while respecting the admissible tire-force and actuator limits. This motivates the use of Model Predictive Control (MPC), which embeds vehicle dynamics in a receding-horizon constrained optimisation and computes, at each step, the control sequence that best balances path-tracking performance, sideslip regulation, and speed objectives. This capability has been demonstrated using both linear MPC \cite{13Hu2022} and NMPC formulations \cite{10Goh2024, 28Siampis2018}, illustrating real-time performance and effective path-tracking during drifting maneuvers.

%Despite this progress in autonomous drifting, the 
Assisted drifting, where a human driver remains in control and the controller acts as a co-pilot, is a qualitatively different problem that has received comparatively little attention. Rather than tracking a fully prescribed reference trajectory, an assist system must infer driver intent in real time and amplify it through a bounded intervention that feels intuitive and transparent to the driver. Sun et al. developed the current state of the art in this domain across a series of papers. In \cite{17Sun2025}, the concept of sideslip rate restriction was established as the key design lever: by limiting how rapidly sideslip grows, the system preserves the driver's reaction time for counter-steering without dictating the exact vehicle attitude. In \cite{15Sun2025}, the framework was extended to a full rally track via a four-mode state machine, Normal, Drift, Sideslip Recovery, and Counter-Steer Recovery, using a Trajectory Radius Predictor to estimate path curvature online and generate the yaw rate reference without any prior knowledge of the path geometry. Drift intention is detected from driver inputs and vehicle motion signals \cite{16Sun2025}, and a torque vectoring controller restricts sideslip rate in Drift Mode while driving sideslip back to zero in Recovery Mode.  The framework in \cite{15Sun2025} nonetheless leaves several open issues. The trajectory-radius predictor is limited to hairpin corners, actuator constraints are not explicitly handled, and the framework assumes driver steering actions that are difficult for non-expert drivers. More fundamentally, the method is developed for a four-wheel-drive electric powertrain with independently controlled wheel torques. This rich actuation architecture is also what allows the controller to avoid explicitly precomputing a drift equilibrium.

These observations motivate revisiting the assisted drifting problem under a different and more restrictive actuation setting. %In particular, the present work considers 
A rear-wheel-drive vehicle is here considered, where the available control authority is limited to steering and rear-drive torque. % and the simplifications enabled by per-wheel torque vectoring no longer apply. 
The goal is to reformulate the assisted-drifting problem for this reduced-actuation configuration and to develop a control architecture that remains intuitive for the driver while explicitly accounting for vehicle and actuator limitations.

To this end, we propose an NMPC framework for assisted vehicle drifting. When the controller is activated, the vehicle speed reference is frozen at its value at the activation instant, while the sideslip reference is commanded by the driver through the steering wheel. During controller operation, the steering and throttle commands are decoupled from the wheel and motor actuators through steer-by-wire and drive-by-wire interfaces, so that the driver specifies the desired sideslip evolution while the NMPC computes the steering and rear-drive actions required to achieve it. Controller activation and deactivation are handled by a dedicated supervisory module, which engages the controller only when the vehicle is in a sharp turn and the driver fully depresses the throttle pedal, thereby ensuring deliberate initiation.
The proposed framework is built around two main elements. The first is a reduced-order NMPC formulation showing that a simple single-track prediction model with a basic tire representation is sufficient to stabilize a high-fidelity simulated RWD vehicle in drifting, using only steering and throttle as control inputs and requiring only vehicle speed and sideslip as references. The second is a non-invasive activation architecture based on phase-plane analysis and driver-intention detection, designed to leave normal driving unaffected while enabling intuitive assisted drift initiation. The overall framework is validated in a high-fidelity simulation environment, where drifting maneuvers are performed with sideslip references continuously adjusted by the driver through the steering wheel, confirming the effectiveness of the proposed approach.
%%%%%%%%%%%%%%%%%%%%%%%%%%%%%%%%%%%%%%%%%%%%%%%%%%%%%%%%%%%%%%%%%%
\section{Vehicle Model}
\label{sec:vehicle_mode }
For the MPC vehicle model, a nonlinear single-track model is employed. Although it is one of the simplest planar vehicle models, it has been shown to effectively capture key aspects of vehicle handling behaviour \cite{3Milani2021}. Several studies have also demonstrated that the single-track model is suitable for analysing highly nonlinear maneuvers such as drifting \cite{1Hindiyeh2013}. %To maintain low system complexity, neither lateral nor longitudinal load transfer is considered, and each axle is represented by a single tire—one at the front and one at the rear. %Additionally, rolling and pitching motions of the chassis are neglected. 
For a rear-wheel drive vehicle, the dynamics of the model may be expressed by the following equations of motion \cite{12Goel2020}
\begin{subequations}\label{eq:stm}
\begin{align}
\dot{V} = \frac{1}{m}[&-F_{yf}\sin(\delta-\beta)\nonumber\\
&+F_{xr}\cos(\beta)+F_{yr}\sin(\beta)],\label{eq:V_dot}\\
\dot{\beta} = \frac{1}{mV}[&+F_{yf}\cos(\delta-\beta)\nonumber\\
&- F_{xr}\sin(\beta)+F_{yr}\cos(\beta)]-r,\label{eq:beta_dot}\\
\dot{r}=\frac{1}{I_{zz}}[&a(F_{yf}\cos(\delta)) -bF_{yr}],\label{eq:r_dot}
\end{align}
\end{subequations}
where $F_{xr}$ denotes the rear longitudinal force, $F_{yf}$ and $F_{yr}$ represent the front and rear lateral tire forces, respectively. Furthermore, $r$ denotes the yaw rate, $V$ the vehicle total velocity, and $\beta$ the sideslip angle evaluated at the vehicle Centre Of Gravity (COG). The variable $\delta$ represents the steering angle at the front wheels. Finally, $a$ and $b$ denote the distances from the COG to the front and rear axles, respectively.
Fig. \ref{fig:STM_model} shows a schematic representation of the vehicle model.

\begingroup

\begin{figure}[!ht]
    \centering
    \begin{adjustbox}{max width=.8\linewidth}

    \begin{tikzpicture}[>=latex, remember picture, node distance=\versep and \horsep]
    >=Stealth,
    thick,
    font=\large
]

\pgfmathsetmacro{\axisAngle}{40}
\pgfmathsetmacro{\adist}{2.5}
\pgfmathsetmacro{\bdist}{3}
\pgfmathsetmacro{\ww}{0.45}
\pgfmathsetmacro{\wl}{1.1}
\pgfmathsetmacro{\deltaAngle}{30}
\pgfmathsetmacro{\betaAngle}{34}

% Derived positions
\pgfmathsetmacro{\fx}{\adist * cos(\axisAngle)}
\pgfmathsetmacro{\fy}{\adist * sin(\axisAngle)}
\pgfmathsetmacro{\rx}{-\bdist * cos(\axisAngle)}
\pgfmathsetmacro{\ry}{-\bdist * sin(\axisAngle)}
\pgfmathsetmacro{\frontAngle}{\axisAngle + \deltaAngle}
\pgfmathsetmacro{\frontAngle}{\axisAngle + \deltaAngle}
\pgfmathsetmacro{\alphaFAngle}{\frontAngle-15}
\pgfmathsetmacro{\alphaRAngle}{\axisAngle-15}

% Blue body line
\draw[darkgray, line width=1.95pt] (\rx,\ry) -- (\fx,\fy);

% Rear wheel
\begin{scope}[shift={(\rx,\ry)}, rotate=\axisAngle]
  \fill[teal!20, opacity=0.9] (-\wl,-\ww) rectangle (\wl,\ww);
  \draw[black, line width=1pt] (-\wl,-\ww) rectangle (\wl,\ww);
\end{scope}

% Front wheel (steered by delta)
\begin{scope}[shift={(\fx,\fy)}, rotate=\frontAngle]
  \fill[teal!20, opacity=0.9] (-\wl,-\ww) rectangle (\wl,\ww);
  \draw[black, line width=1pt] (-\wl,-\ww) rectangle (\wl,\ww);
\end{scope}

% CoM
\fill[black] (0,0) circle (0.18);
\draw[black, line width=1pt] (0,0) circle (0.18);

\fill[black] (\fx,\fy) circle (0.1);
\fill[black] (\rx,\ry) circle (0.1);

% Distance labels a and b (offset perpendicular)
\pgfmathsetmacro{\perpOff}{1.5}
\pgfmathsetmacro{\perpAngle}{\axisAngle - 90}
\pgfmathsetmacro{\offx}{\perpOff * cos(\perpAngle)}
\pgfmathsetmacro{\offy}{\perpOff * sin(\perpAngle)}

\draw[{Stealth}-{Stealth}, line width=1.2pt]
  (\offx, \offy) -- ({\rx + \offx}, {\ry + \offy})
  node[midway, below right, yshift=-2pt] {$b$};

\draw[{Stealth}-{Stealth}, line width=1.2pt]
  (\offx, \offy) -- ({\fx + \offx}, {\fy + \offy})
  node[midway, below right, yshift=-2pt] {$a$};

% ---- Velocity vector V at CoM ----
\pgfmathsetmacro{\Vangle}{\axisAngle + \betaAngle}
\draw[->, blue!80, line width=1.5pt] (0,0) -- 
  ({1.5*cos(\Vangle)},{1.5*sin(\Vangle)}) node[above, black] {$V$};

% beta arc
\draw[->, black, line width=1.2pt] ({0.85*cos(\axisAngle)},{0.85*sin(\axisAngle)})
  arc[start angle=\axisAngle, end angle=\Vangle, radius=0.95];
\pgfmathsetmacro{\betaMid}{\axisAngle + \betaAngle/2}
\node at ({1.2*cos(\betaMid)},{1.2*sin(\betaMid)}) {$\beta$};

% Yaw rate r
\draw[->, black, line width=1.2pt] (-0.4, -0.7) 
  arc[start angle=-120, end angle=15, radius=0.8];
\node at (0.85, -0.75) {$r$};

% ---- FRONT WHEEL FORCES ----
\pgfmathsetmacro{\frontNorm}{\frontAngle}
\pgfmathsetmacro{\FyfAngle}{\frontAngle + 90}

% Fyf force arrow
\draw[->, orange, line width=2pt] (\fx,\fy) --
  ({\fx + 2.5*cos(\FyfAngle)},{\fy + 2.5*sin(\FyfAngle)})
  node[above left] {$F_{yf}$};

% dashed reference lines at front wheel
\pgfmathsetmacro{\axisNormFx}{\fx + 2.6*cos(\axisAngle)}
\pgfmathsetmacro{\axisNormFy}{\fy + 2.6*sin(\axisAngle)}
\pgfmathsetmacro{\frontNormFx}{\fx + 2.6*cos(\frontNorm)}
\pgfmathsetmacro{\frontNormFy}{\fy + 2.6*sin(\frontNorm)}
\draw[dashed, black] (\fx,\fy) -- (\axisNormFx, \axisNormFy);
\draw[dashed, black] (\fx,\fy) -- (\frontNormFx, \frontNormFy);

% delta arc
\draw[->, thin] ({\fx + 2.3*cos(\axisAngle)},{\fy + 2.3*sin(\axisAngle)})
  arc[start angle=\axisAngle, end angle=\frontNorm, radius=2.3];
\pgfmathsetmacro{\deltaMid}{\axisAngle + \deltaAngle/2}
\node at ({\fx + 2.6*cos(\deltaMid)},{\fy + 2.6*sin(\deltaMid)}) {$\delta$};

% alpha_f arc
\draw[->, black, line width=1pt] (\fx,\fy) --
  ({\fx + 1.8*cos(\alphaFAngle)},{\fy + 1.8*sin(\alphaFAngle)})
  node[above left] {};
\draw[->, thin] ({\fx + 1.5*cos(\frontAngle)},{\fy + 1.5*sin(\frontAngle)})
  arc[start angle=\frontNorm, end angle=\alphaFAngle, radius=1.5];
\pgfmathsetmacro{\alphaFMid}{\alphaFAngle+5}
\node at ({\fx + 1.9*cos(\alphaFMid)},{\fy + 1.9*sin(\alphaFMid)}) {$\alpha_f$};

% ---- REAR WHEEL FORCES ----
\pgfmathsetmacro{\rearNorm}{\axisAngle + 90}

% Fyr force arrow
\draw[->, orange, line width=2pt] (\rx,\ry) --
  ({\rx + 1.7*cos(\rearNorm)},{\ry + 1.7*sin(\rearNorm)})
  node[above left] {$F_{yr}$};

% % dashed reference at rear wheel
% \draw[dashed, gray] (\rx,\ry) -- ({\rx + 1.4*cos(\rearNorm)},{\ry + 1.4*sin(\rearNorm)});

% alpha_r arc
\draw[->, black, line width=1pt] (\rx,\ry) --
  ({\rx + 1.8*cos(\alphaRAngle)},{\ry + 1.8*sin(\alphaRAngle)})
  node[above left] {};
\draw[->, thin] ({\rx + 1.5*cos(\axisAngle)},{\ry + 1.5*sin(\axisAngle)})
  arc[start angle=\axisAngle, end angle=\alphaRAngle, radius=1.5];
\pgfmathsetmacro{\alphaRMid}{\alphaRAngle+5}
\node at ({\rx + 1.9*cos(\alphaRMid)},{\ry + 1.9*sin(\alphaRMid)}) {$\alpha_r$};

% Fxr force arrow (longitudinal, red)
\draw[->, red, line width=2pt] (\rx,\ry) --
  ({\rx + 1.9*cos(\axisAngle)},{\ry + 1.9*sin(\axisAngle)})
  node[above left, xshift=2pt] {$F_{xr}$};

\end{tikzpicture}
\end{adjustbox}

\caption{Schematic representation of the single track model adopted in the paper - the large black dot represents the CoG. %$\alpha_f$ and $\alpha_r$ are the front and rear tire slip angles.
}
\label{fig:STM_model}
\end{figure}
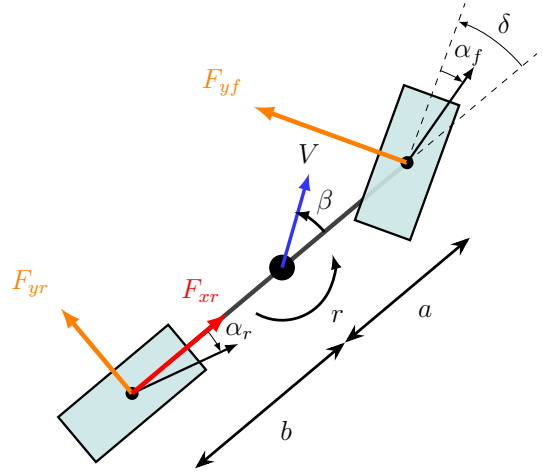
    
\endgroup
% \begin{figure}
%     \centering
%     \includegraphics[width=0.3\textwidth]{Figures/STM_model.eps}
%     \caption{Schematic representation of the single track model adopted in the paper. $\alpha_f$ and $\alpha_r$ are the front and rear tire slip angles.}
%     \label{fig:STM_model}
% \end{figure}
Lateral forces are modelled using a Fiala tire model \cite{8Peterson2023} which, for a generic tire, reads
\begin{equation} \label{eq:Fiala}
\begin{cases}
\begin{aligned}
F_y =\ &C_{\alpha}\tan(\alpha)-\frac{C_{\alpha}^2}{3F_{y,\max}}|\tan(\alpha)|\tan(\alpha) \\
&+\frac{C_{\alpha}^3}{27F_{y,\max}^2}\tan^3(\alpha)
\end{aligned}
& |\alpha|\leq{}\alpha_s \\
F_y=F_{y,\max}\mathrm{sgn}(\alpha) & \text{otherwise}
\end{cases}
\end{equation}

% \begin{equation} \label{eq:tanh}
% F_y=F_{y,\max}\tanh\left(\frac{\alpha}{\alpha_s} \right)
% \end{equation}
% %
% where $\alpha_s$ is the saturation slip angle
% \begin{equation}
%     \alpha_s = \arctan\left(\frac{F_{y,\max}}{C_\alpha}\right)
% \end{equation}
where $C_\alpha$ is the axle cornering stiffness, $\alpha_s$ is the saturation slip angle %(the sum of the two tires) 
and $F_{y,\max}$ is the maximum available axle lateral force, calculated using the friction circle \cite{8Peterson2023}%\RDC{ADD A REFERNCE}:
\begin{equation}\label{eq:FY_max}
F_{y,\max}=\sqrt{(\mu F_z)^2-F_{xr}^2},
\end{equation}
where $F_z$ is the vertical load and $\mu$ the friction coefficient.

Kinematic equations define front and rear tire slip angles
\begin{align}\label{eq:alpha}
\alpha_f &=\delta-\arctan\biggl(\frac{V\sin(\beta)+ar}{V\cos(\beta)}\biggr),\\
\alpha_r &= -\arctan\biggl(\frac{V\sin(\beta)-br}{V\cos(\beta)}\biggr).
\end{align}
%
% while $\alpha_s$ is the maximum slip angle beyond which the tire starts to slide
% %
% \begin{equation}\label{eq:alpha_slide}
% \alpha_s=\arctan\biggl(\frac{3F_{y,max}}{C_{\alpha}}\biggr)
% \end{equation}
% %
The vehicle parameters are reported in Table \ref{tab:param}.

\begin{table}
    \centering
        \caption{Vehicle and tire parameters.}
    \begin{tabular}{|l|c|c|}
        \hline
        \textbf{Parameter} & \textbf{Symbol} & \textbf{Value}\\
        \hline
        Mass & $m$ & 1937\,kg \\
        Yaw moment of inertia & $I_{zz}$ & 1287\,kg\,m$^2$ \\
        COG to front axle distance & $a$ &  1.347\,m \\
        COG to rear axle distance & $b$ & 1.393\,m \\
        Front cornering stiffness & $C_{\alpha f}$ & 132\,kN/rad \\
        Rear cornering stiffness & $C_{\alpha r}$ & 158\,kN/rad\\
        Tire-road friction coefficient & $\mu$ & 1  \\
        Steering ratio & $\kappa_\delta$ & 0.048\\
        \hline
    \end{tabular}

    \label{tab:param}
\end{table}

\section{Assisted Drifting Framework}

The overall architecture of the proposed framework is illustrated in Fig.~\ref{fig:block-scheme}. The driver provides throttle and steering wheel inputs, denoted by $\tau_d$ and $\delta_d$, respectively. When a drifting intention is detected, the driver inputs are decoupled from the vehicle actuators. In this phase, the steering wheel input is interpreted as a sideslip reference for the NMPC controller, which then assumes full authority over both longitudinal and lateral vehicle control. Once the deactivation conditions are satisfied, full control authority is smoothly returned to the driver.
In this section, we detail the activation and deactivation strategies, as well as the NMPC-based control strategy employed for drifting.
\begingroup
\begin{figure}[!ht]
    \centering
    \begin{tikzpicture}[
        >=latex,
        remember picture,
        node distance=\versep and \horsep,
        block/.style={
            draw=black!60,
            thick,
            rounded corners=3pt,
            align=center,
            font=\small\scshape,
            inner sep=6pt
        },
        line/.style={
            ->,
            thick,
            draw=black!75
        }
    ]

        % Driver
        \node (dri) [
            block,
            fill=orange!20,
            minimum height=.45\ieeecolw,
            minimum width=.1\ieeecolw
        ] {\rotatebox{90}{Driver}};

        % NMPC
        \node (mpc) [
            block,
            fill=teal!20,
            right=.4\horsep of dri.south east,
            anchor=south west
        ] {NMPC Controller};

        % AD Strategy
        \node (ads) [
            block,
            fill=teal!20,
            above right=.5\versep and .5\horsep of mpc.north,
            anchor=south
        ] {A--D Strategy};

        \coordinate (ads-a) at ($(ads.north)+(-10pt,0)$);
        \coordinate (ads-b) at ($(ads.north)+(10pt,0)$);

        % VI-CRT block
        \node (crt) [
            right=3\horsep of dri.north east,
            anchor=north west,
            inner sep=0pt,
            label={[anchor=south east]south east:
            \textcolor{red!70!black}{\footnotesize\scshape VI-CRT}}
        ] {\includegraphics[height=.25\ieeecolw]{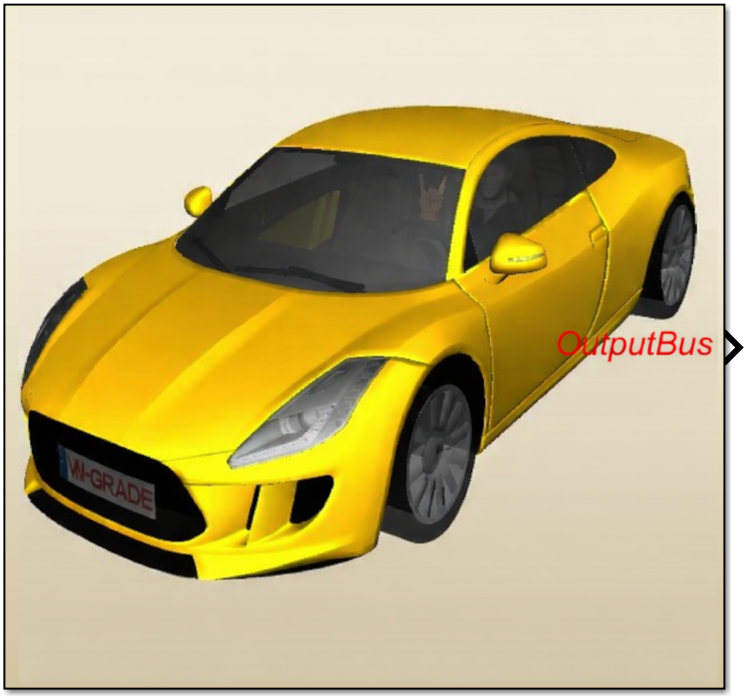}};

        % Integrator
        \node (int) [
            block,
            fill=gray!15,
            left=.75\horsep of crt
        ] {$\int$};

        % Coordinates
        \coordinate (crt-a) at ($(crt.west)+(0,10pt)$);
        \coordinate (crt-b) at ($(crt.west)+(0,-20pt)$);
        \coordinate (mpc-a) at ($(mpc.north)+(-30pt,0)$);
        \coordinate (c1) at ($(crt.east)+(10pt,0)$);
        \coordinate (c3) at ($(dri.east)+(0,45pt)$);

        % Connections
        \draw[line] (c3) --
            node[above]{\scriptsize$\delta_d,\tau_d$}
            (c3-|crt.west);

        \draw[line] (int) --
            node[above]{\scriptsize$\delta_c,\tau_c$}
            (int-|crt.west);

        \draw[line] (mpc-a) |-
            node[above,near end]{\scriptsize$u_\delta,u_\tau$}
            (int);

        \draw[line] (ads-b) |- (crt-b);

        \draw[line] (ads.north-|int) -- (int.south);

        \draw[line] (crt.east) -- (c1) |-
            node[above,near end]{\scriptsize$V,\beta,r,\tau_d$}
            (ads.east);

        \draw[line] (crt.east) -- (c1) |-
            node[above,near end]{\scriptsize$V,\beta,r,\delta_d$}
            (mpc.east);

    \end{tikzpicture}
    \caption{Architecture of the proposed control framework validated in VI-CarRealTime.}
    \label{fig:block-scheme}
\end{figure}
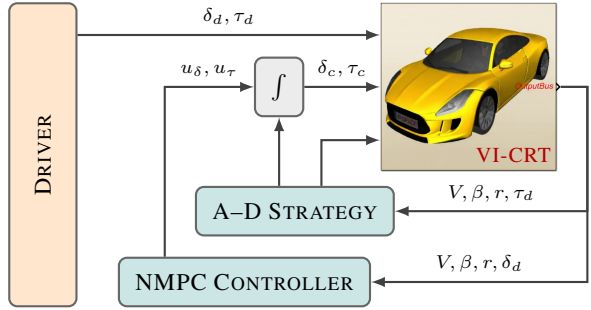
\endgroup % clearly better

\subsection{Activation and deactivation strategies}
The interaction design between the driver and the assisted drifting controller is a key element of the overall system, as it directly influences usability, controllability, and driver acceptance in closed-loop operation. In particular, the activation phase represents the most critical aspect of this interaction. Unlike conventional stability controllers that intervene reactively for safety, an assisted drifting controller must engage only when an intention to drift is detected in the driver behaviour.
Drift initiation is typically characterized by the generation of high lateral acceleration followed by a rapid increase in yaw moment to induce rear tire saturation and large sideslip angles. This is commonly achieved through aggressive steering combined with throttle application (power oversteer) or via transient handbrake input \cite{gillespie1992fundamentals, 5Velenis2010}. Once these actions push the vehicle into a strongly nonlinear regime, a counter-steering maneuver becomes necessary to prevent the vehicle from spinning out \cite{Edelmann2011}. This maneuver can typically be performed successfully only by experienced drivers. For this reason, this task is delegated to the controller, that should activate promptly - before counter-steering is required.
%counter-steer as required.

To formalize this concept, % it is necessary to analyze the phase-plane during a drift maneuver. 
Righetti \cite{righetti2024drifting} shows that a saddle-node bifurcation is observed in the phase-plane when counter-steering action is required. Furthermore, an additional indicator of this phenomenon is the yaw rate approaching its maximum steady-state value, $r_{ss,\max}$ \cite{BobierTiu2019}, defined as

\begin{equation}
    r_{ss,\max} =  \frac{a_{y,ss,\max}}{V\cos(\beta)}
\end{equation}
where $a_{y,ss,\max}$ is the maximum lateral acceleration at steady-state
\begin{equation}
    a_{y,ss,\max} = \min \left(\mu g, \frac{a+b}{am}\sqrt{\left(\mu F^{ST}_{z,r} \right)^2 - F_{xr}^2} \right)
\end{equation}
where $F^{ST}_{z,r}$ is the static vertical load on the rear axis.
From this, an activation logic is formalized
\begin{equation}
    r > \lambda r_{ss,\max}, \qquad \tau_{d} = 100 
\end{equation}
where $\lambda \in [0,1]$ is a scaling coefficient introduced to tune the activation sensitivity of the controller. In particular, it regulates how easily the driver can trigger controller engagement. As $\lambda$ increases, the driver must execute a turn with progressively higher lateral acceleration before applying full throttle to activate the controller. The requirement of full throttle acts as an intentional activation trigger, ensuring that controller engagement results from a deliberate driver action rather than from only operating conditions. This heuristic averts unintended controller interventions% that the controller enters when it is not intended
, for example when driving on a straight line with full throttle. This heuristic also ensures that the controller is activated only when the vehicle is operating near a loss-of-control condition and a full-throttle input is applied.
The deactivation phase is comparatively simpler. To safely return control of the steering and throttle to the driver, the controller requires the driver to command a zero sideslip reference via the steering input while fully releasing the throttle pedal. Once the vehicle has stabilized in straight-line driving conditions for a duration of $\Delta t$, the controller is smoothly deactivated, thereby restoring full control authority to the driver.

\begin{equation}
    \delta_d(t) = 0, \quad \tau_{d}(t) = 0, \quad \beta(t) \approx 0 \quad \text{for } t \in [t^*, t^*+\Delta t].
\end{equation}

\subsection{NMPC control for assisted drifting}
In this work, a NMPC scheme is developed to generate steering and throttle inputs in a receding-horizon approach in order to achieve and sustain a drifting maneuver. The NMPC formulation is based on an input velocity-form representation of the dynamic bicycle model \eqref{eq:stm}, where the state is a combination of measured vehicle states and previous input states. Specifically, the state vector is
\begin{equation}
\xi = \begin{bmatrix} V & \beta & r & \delta_c & \tau_c \end{bmatrix}^\top,
\end{equation}
the system dynamics can be written as
\begin{align}\label{eq:mpc-vel-form}
    \dot\xi &= f(\xi,u) = 
    \left[\begin{array}{c|c}
          \eqref{eq:V_dot} \cdots \eqref{eq:r_dot} &  u^\top
    \end{array}\right]^\top
\end{align}
where time derivatives of the physical inputs are treated as control variables $u = [\udelta, \ufxr]^\top$. This reformulation promotes smoother actuator behaviour and does not require explicit computation of input references at the specific drift equilibrium configuration.
Based on this model, the following continuous-time OCP can be formulated
\begin{subequations}\label{eq:mpc-ocp-cont}
\begin{align}
    \min_{\xi,u} \quad & \int_0^{T} \ell(\xi(t),u(t)) ~\mathrm{d}t + M(\xi(T)) \label{eq:mpc-ocp-cont-a}\\
    \mathrm{s.t.}\quad & \xi(0) = \overline\xi, \label{eq:mpc-ocp-cont-b}\\
    &\dot\xi(t) = f(\xi(t),u(t)), \quad t\in[0,T] \label{eq:mpc-ocp-cont-c}\\
    &\underline{r}(t) \leq r(\xi(t),u(t)) \leq \overline{r}(t) \quad t\in[0,T] \label{eq:mpc-ocp-cont-d}
\end{align}
\end{subequations}
To reach the reference velocity and sideslip angle effectively, stage and terminal cost functions in \eqref{eq:mpc-ocp-cont-a} are defined as
\begin{subequations}\label{eq:mpc-costs}
\begin{align}
\ell(\xi,u) &= \frac{1}{2}\| h(\xi,u) \|_W \quad M(\xi) = \frac{1}{2}\| h_N(\xi,u) \|_{W_N}\\
h(\xi,u) 
&= 
\begin{bmatrix}
e_V 
&e_\beta 
% &\delta 
% &F_{xr}
&\udelta 
&\ufxr
\end{bmatrix}^\top,\label{eq:stage_cost}\\
h_N(\xi) &= \begin{bmatrix} e_V & e_\beta \end{bmatrix}^\top,
\end{align}
\end{subequations}
where velocity and sideslip tracking errors are defined as
\begin{align}
e_V &= V- V^{\mathrm{ref}},
&e_\beta &= \beta - \beta^{\mathrm{ref}}.
\end{align}
In this framework, the NMPC controller effectively operates as an online equilibrium solver. Referring to Fig.~\ref{fig:block-scheme}, once the drifting intention is detected, the reference velocity $V^{\mathrm{ref}}$ is set equal to the vehicle velocity at the triggering instant. Simultaneously, the reference sideslip angle $\beta^{\mathrm{ref}}$ is obtained from the steering wheel input through a linear mapping defined~as
\begin{equation}
\beta^{\mathrm{ref}}(t) = \kappa_{\beta}\,\delta_d(t).
\end{equation}
where $\kappa_\beta$ is a scaling factor.
One of the main advantages of NMPC is the ability to explicitly enforce constraints on both states and inputs. In the proposed approach, the constraints in \eqref{eq:mpc-ocp-cont-d} are imposed only on the actuators and their slew rates
\begin{align}
r_k &=
\begin{bmatrix}
\delta_c & \tau_c & \udelta & \ufxr
\end{bmatrix}^\top,
&r_N &=
\begin{bmatrix}
\delta_c & \tau_c
\end{bmatrix}^\top.
\label{eq:rkN}
\end{align}

To solve the OCP~\eqref{eq:mpc-ocp-cont}, a direct optimal control method is employed. In particular, \eqref{eq:mpc-ocp-cont} is discretized into $N = T/T_s$ shooting intervals using a multiple-shooting scheme, where $T_s$ represents the sampling time \cite{Diehl2009}
\begin{subequations}\label{eq:mpc-ocp-disc}
\begin{align}
\min_{\xi_k, u_k} \quad & \sum_{k=0}^{N-1} \ell(\xi_k,u_k) \cdot T_s + M(\xi_N) \label{eq:mpc-ocp-disc-a}\\
\text{s.t.} \quad & \xi_0 = \overline\xi, \label{eq:mpc-ocp-disc-b}\\
& \xi_{k+1} = \Phi_k(\xi_k, u_k), \quad k = 0, \dots, N-1, \label{eq:mpc-ocp-disc-c}
\\
& \underline{r}_k \le r_k(\xi_k, u_k) \le \bar r_k, \quad k = 0, \dots, N-1, \label{eq:mpc-ocp-disc-d}
\\
& \underline{r}_N \le r_N(\xi_N) \le \bar r_N. \label{eq:mpc-ocp-disc-e}
\end{align}
\end{subequations}
The resulting discrete-time OCP~\eqref{eq:mpc-ocp-disc} is solved in real time using the \textsc{MATMPC} toolbox~\cite{Matmpc}. At each sampling instant, one Sequential Quadratic Programming (SQP) iteration is performed according to the Real-Time Iteration (RTI) scheme, employing an explicit fourth-order Runge-Kutta (ERK4) integrator \cite{Diehl2009}.
The first element of the optimal control sequence $u_0^{*}$ is applied to the vehicle, and the optimization procedure is repeated in a receding horizon fashion using updated state measurements. Once activated, the computed inputs are integrated and transmitted to the vehicle. Owing to the steer-by-wire architecture, the effective steering wheel angle is defined as the controller steering command when the controller is active, and as the driver steering input otherwise.
\begin{equation}
\delta_w=  
\begin{cases}
    \delta_c, \qquad \text{if controller is active} \\
    \delta_d, \qquad \text{otherwise.}
\end{cases}
\end{equation}
The steering angle at the wheel can then be computed using the steering ratio $\kappa_\delta$
\begin{equation}
    \delta = \kappa_\delta \delta_w
\end{equation}

% The state constraints in \eqref{eq:NLP_states_constr} are defined by
% \[
% r_k(\xi_k, u_k) : \mathbb{R}^{n_x} \times \mathbb{R}^{n_u} \to \mathbb{R}^{n_r},
% \]
% and the terminal constraint in \eqref{eq:NLP_terminal} by
% \[
% r_N(\xi_N) : \mathbb{R}^{n_x} \to \mathbb{R}^{n_r}.
% \]

\section{Simulation setup and results}
The validation procedure is structured into two phases. The first phase aims to demonstrate that, under the assumption of perfect knowledge of the vehicle model, the proposed NMPC formulation guarantees convergence to a drift equilibrium point even when the reference supplied to the controller comprises only a subset of the system states.
The second phase focuses on validation within a high-fidelity vehicle simulation environment \texttt{VI-CarRealTime} (VI-CRT), where the controller performance is assessed under more realistic operating conditions. First, it is shown that the NMPC can successfully initiate and sustain a stable drift maneuver, driving the system to an equilibrium. %Subsequently, 
It is then demonstrated that, given physically plausible steering wheel and throttle inputs from the driver, the controller activates smoothly according to the defined activation logic, steers the vehicle toward the desired drifting condition, and dynamically regulates the maneuver based on the sideslip angle reference implicitly commanded through steering inputs. Finally, the controller ensures a seamless transition back to nominal straight-line driving, progressively restoring full control authority to the driver.
The analyses are performed on a fully electric vehicle equipped with one electric motor delivering the driving torque to the rear wheels through an open differential.

The simulation frequency is set to $1$ kHz, while the controller frequency is set to $100$ Hz. Let $\lambda = 0.9$, $\delta_{c}^{\max} = 3\pi$, $\tau_c^{\max} = 100$, $u_{\delta_c}^{\max} = \pi$, $u_{\tau_c}^{\max} = 1500$, $V^{\max}=50\,\mathrm{m}/\mathrm{s}$, $\beta^{\max}=\pi/3$, $\kappa_\beta = -\frac{1}{5}$, the NMPC parameters are% set as following:
\begin{equation}
\begin{aligned}
    \overline{r}_k &=
    \left[\delta_{c}^{\max}, \tau_c^{\max}, u_{\delta_c}^{\max}, u_{\tau_c}^{\max}\right]^\top,\\
    \underline{r}_k &=
    \left[-\delta_{c}^{\max}, 0, -u_{\delta_c}^{\max}, -u_{\tau_c}^{\max}\right]^\top,\\
    \overline{r}_N &=\left[\delta_{c}^{\max}, \tau_c^{\max}
    \right]^\top,\\
    \underline{r}_N &=\left[-\delta_{c}^{\max}, -\tau_c^{\max}
    \right]^\top,\\
    W &= \mathrm{diag}\left( \frac{10^3}{V^{\max}}, \frac{10^3}{\beta^{\max}}, \frac{15\cdot10^2}{u_{\delta_c}^{\max}}, \frac{10^3}{u_{\tau_c}^{\max}}\right),\\
    W_N &= \mathrm{diag}\left( \frac{10^3}{V^{\max}}, \frac{10^3}{\beta^{\max}}\right).
\end{aligned}
\end{equation}

\subsection{Equilibrium convergence under nominal conditions}
A preliminary analysis aims to experimentally demonstrate that, under nominal conditions, the proposed NMPC formulation can converge to a drift equilibrium without explicit pre-computation, relying solely on velocity and sideslip angle references. To this end, an arbitrary set of reference values are assigned to the longitudinal velocity $V^{\mathrm{ref}}$ and the sideslip angle $\beta^{\mathrm{ref}}$. Consequently, the drift equilibrium point corresponding to this reference velocity and sideslip angle is computed offline
\begin{equation}
    x^{\mathrm{eq}} = \begin{bmatrix}
        V^{\mathrm{eq}} & \beta^{\mathrm{eq}} & r^{\mathrm{eq}} & \delta_w^{\mathrm{eq}} & \tau^{\mathrm{eq}} 
    \end{bmatrix}^\top
\end{equation}
where $V^{\mathrm{eq}} = V^{\mathrm{ref}} = 20 \ \text{m/s}$ and $\beta^{\mathrm{eq}} = \beta^{\mathrm{ref}} = -0.43 \ \text{rad}$ ($-25\  \text{deg}$). Multiple initial conditions are considered to challenge the controller to achieve the prescribed velocity and sideslip angle while steering the system toward the corresponding drift equilibrium.
% The initial condition is set to $\xi_0 = \left[15, 0, 0, 0, 0 \right]^\top$, corresponding to straight-line driving at an initial velocity of $V_0 = 54 \ \text{km/h}$, thereby challenging the controller to simultaneously achieve the prescribed velocity and sideslip angle references. 

\begin{figure}
    \centering
    \includegraphics[width=0.5\textwidth]{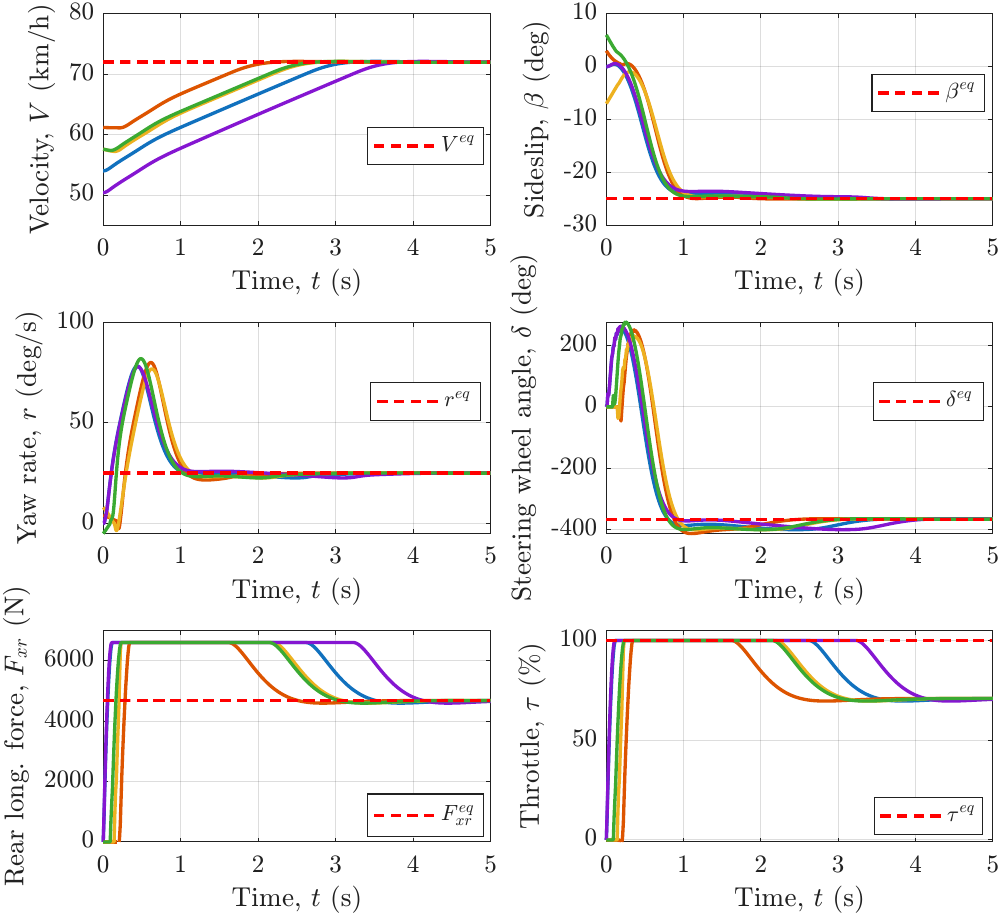}
    \caption{Validation of the controller using the single-track model under nominal conditions starting from different initial conditions.}
    \label{fig:sim_plot}
\end{figure}

Fig. \ref{fig:sim_plot} shows that the NMPC successfully drives the system toward the corresponding drift equilibrium from all considered initial conditions. This result confirms that, when the prediction model accurately represents the system dynamics, the proposed control formulation is capable of autonomously identifying and stabilizing the steady-state drifting condition associated with the prescribed references.
\subsection{Controller validation in high-fidelity simulation}
\begin{figure}
    \centering
    \includegraphics[width=0.5\textwidth]{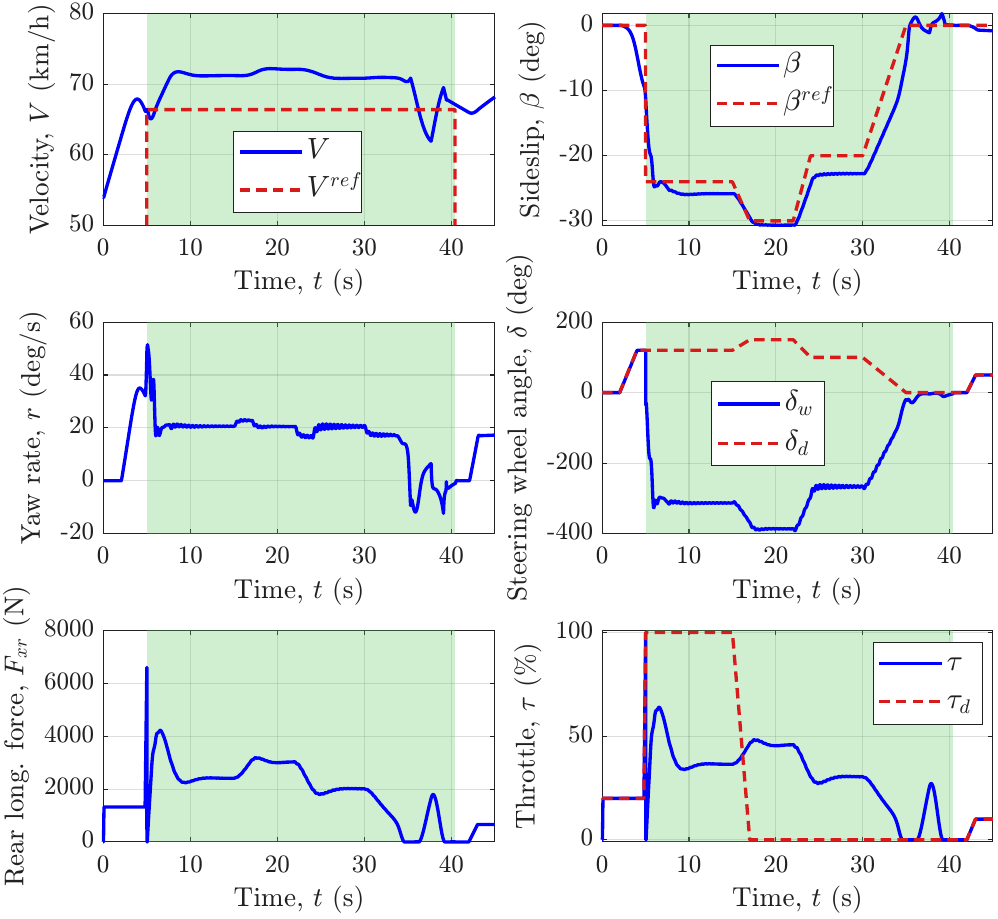}
    \caption{Validation of the controller in the VI-CarRealTime simulation environment. The green shaded region indicates the time interval during which the controller is active.}
    \label{fig:crt_plot}
\end{figure}

To validate the activation logic and the overall architecture, the test shown in Fig. \ref{fig:crt_plot} is conducted, where the steering and throttle profiles are generated to mimic the behavior of a human driver.
% It is interesting to note that lateral slip is accurately tracked well before convergence is reached.

From $t = 0 \ s$ to $t = 5 \ s$, the driver applies a steering ramp while maintaining a constant throttle input in order to approach a high lateral-acceleration condition. At $t = 5 \ s$, the driver applies full throttle, thereby triggering the controller activation logic, which decouples the steering wheel and throttle pedal from direct vehicle control. During the interval $5 \ s \leq t < 15 \ s$, the driver maintains a constant sideslip angle reference through the steering wheel, allowing the controller to stabilize the vehicle in a neighbourhood of the velocity and sideslip references. Subsequently, in the interval $15 \ s \leq t < 22 \ s$, the driver gradually increases the sideslip angle reference to a higher value of $\beta$ and then maintains it constant. The controller correspondingly adapts the vehicle dynamics to track the updated reference. In the following phase, $22 \ s \leq t < 30 \ s$, the driver progressively decreases the steering input, inducing a less pronounced drifting condition. At $t = 30 \ s$, the driver begins to reduce the steering angle until it reaches zero. The controller then guides the vehicle back to a stable straight-line driving condition and smoothly deactivates at $t = 40.4 \ s$, thereby restoring full control authority to the driver. From $t = 43 \ s$, the driver resumes steering and throttle inputs. It is important to note that the tracking, especially for the vehicle speed, is not fully accurate. This behaviour can be attributed to significant model mismatch, reflecting the controller’s trade-off between following the references and finding a steady control action that stabilizes the real vehicle. 

\begin{figure}
    \centering
    \begin{subfigure}[b]{0.49\linewidth}
        \centering
        \includegraphics[width=\linewidth]{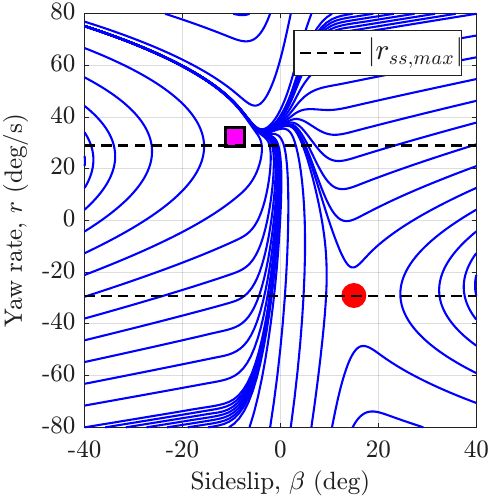}
        \caption{}
        \label{fig:pp_a}
    \end{subfigure}
    \hfill
    \begin{subfigure}[b]{0.49\linewidth}
        \centering
        \includegraphics[width=\linewidth]{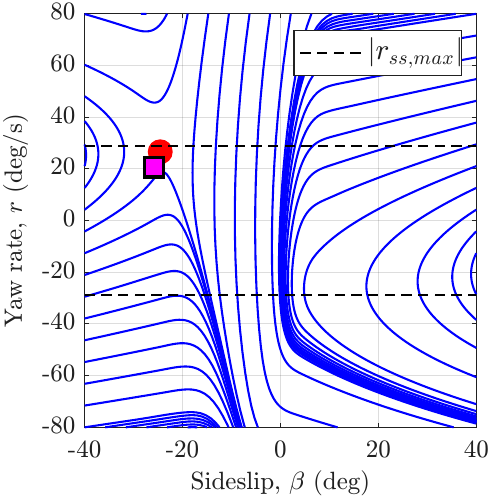}
        \caption{}
        \label{fig:pp_b}
    \end{subfigure}
    \caption{Phase-plane trajectories of the uncontrolled single-track model immediately before controller activation (a) at $t = 4.8 \ s$ and during the drifting maneuver (b) at $t = 13 \ s$. The pink square represents the current vehicle state, while the red dot indicates the corresponding drifting equilibrium.}
    \label{fig:pps}
\end{figure}

As far as the effectiveness of the activation logic is concerned, Fig.~\ref{fig:pp_a} shows that right before controller activation the vehicle yaw rate exceeds $r_{ss,\max}$, indicating that a counter-steering action is necessary to prevent the vehicle from spinning out. This behaviour is further illustrated in Fig. \ref{fig:crt_plot}, where at $t = 5 \ s$ the controller immediately applies a counter-steering input upon activation. It should be noted that the only equilibrium in this region corresponds to an unstable drifting condition, confirming that a saddle-node bifurcation has occurred. Furthermore, this equilibrium is in the opposite quadrant with respect to the current vehicle state, as the counter-steering has not yet been executed \cite{righetti_sae}. Fig.~\ref{fig:pp_b} illustrates the phase-plane trajectories during the drifting maneuver, showing that the vehicle is within a neighbourhood of the nominal equilibrium point.

%%%%%%%%%%%%%%%%%%%%%%%%%%%%%%%%%%%%%%%%%%%%%%%%%%%%%%%%%%%%%%%%%%
\section{Conclusions}
\label{sec:conclusions}
This work presented an assisted drifting control framework based on a simple single-track vehicle model with basic tire dynamics, relying solely on throttle and steering inputs. The main contribution lies in enabling the driver to actively shape the drifting behaviour by dynamically adjusting the sideslip reference through steering wheel, while an intuitive activation logic based on phase-plane analysis ensures controller intervention only during turning conditions and upon conscious driver initiation via full throttle application. High-fidelity simulation results validate the effectiveness of the proposed NMPC-based architecture in stabilizing drifting maneuvers around the requested sideslip reference. Future work will focus on human-in-the-loop validation supported by statistical analyses, as well as on robustness assessments under varying driving conditions.

%%%%%%%%%%%%%%%%%%%%%%%%%%%%%%%%%%%%%%%%%%%%%%%%%%%%%%%%%%%%%%%%%%
\section*{ACKNOWLEDGMENTS}
% Replace with acknowledgments or remove if none
This work was partially supported by the Italian Ministry under the National Recovery and Resilience Plan (PNRR), funded by the European Union – NextGenerationEU.

%%%%%%%%%%%%%%%%%%%%%%%%%%%%%%%%%%%%%%%%%%%%%%%%%%%%%%%%%%%%%%%%%%
%\addtolength{\textheight}{-12cm}
%\vspace{10mm}
\bibliographystyle{IEEEtran}
% Your .bib file here
\bibliography{ref} 
	
\end{document}